\documentstyle[aps,prd,preprint]{revtex}
\begin{document}
\draft
\title{Limits on Active-Sterile Neutrino Mixing and
	the Primordial Deuterium Abundance}
\author{Christian Y. Cardall and George M. Fuller}
\address{
Department of Physics, University of California, San Diego,
La Jolla, California 92093-0319}
\date{\today}
\maketitle
\begin{abstract}
Studies of limits on active-sterile neutrino mixing
derived from big bang nucleosynthesis considerations are 
extended to consider the dependance of these constraints on
the primordial deuterium abundance. This study is motivated by
recent measurements of D/H in quasar absorption systems, which
at present yield discordant results. Limits on active-sterile
mixing are somewhat relaxed for high D/H. For low D/H
($\approx 2 \times 10^{-5}$),
no active-sterile neutrino mixing is allowed by currently 
popular upper
limits on the primordial $^4$He abundance $Y$.  
For such low primordial D/H values, 
the observational inference of 
active-sterile neutrino mixing by upcoming  
solar neutrino experiments would imply that $Y$ has been
systematically underestimated, unless there is new physics
not included in standard BBN.
\end{abstract}


Upper limits on the abundance of $^4$He produced in big bang
nucleosynthesis (BBN) have been used to limit 
mixing between active ($\nu_e$, $\nu_{\mu}$, or $\nu_{\tau}$)
and sterile ($\nu_s$, no standard model interactions) neutrinos
\cite{steriles}. 
In this paper, we point out and discuss how these constraints
are dependant on the adopted primordial deuterium abundance.
Previous limits on sterile neutrino
mixing have assumed a value for the lower bound on
the baryon-to-photon ratio $\eta$ derived from interstellar
medium and solar system measurements of deuterium (D) and 
$^3$He, and models of chemical and galactic evolution. 
Recent measurements of D/H in quasar absorption
systems (QAS) have yielded discordant values of 
this ratio, some
higher than previously derived ranges \cite{hid}, 
and some lower \cite{lowd}. 
Several factors make an investigation of the 
primordial D/H dependance
of BBN constraints on active-sterile neutrino mixing timely:
the discordant QAS measurements of D/H; the fact
that future solar neutrino experiments may be able to 
distinguish and identify $\nu_e-\nu_s$ mixing \cite{nuexp}; 
and the use of sterile 
neutrinos in schemes for neutrino 
masses and mixings that explain all
available data \cite{numod}.

As is well known ({\it e.g.}, Ref. \cite{kolb}), 
the abundance of $^4$He produced 
by BBN is essentially 
determined by the ratio of neutron to proton number densities
(n/p) at ``weak freeze-out'' (WFO). WFO occurs when the reactions
that interchange neutrons and protons proceed too slowly 
relative to the expansion rate of the universe to keep n/p
at its equilibrium value of n/p $\approx \exp (-\Delta m / T)$. Here
$\Delta m \equiv m_n-m_p \approx 1.293$ MeV 
is the neutron-proton mass difference, and $T$ is
the photon temperature. Mixing between active and sterile
neutrinos increases (n/p)$_{\rm{WFO}}$, and therefore the 
primordial $^4$He mass fraction $Y$, in two ways. First,
active-sterile neutrino mixing effectively brings more degrees of 
freedom into thermal contact, increasing the energy
density and hence the expansion rate of the universe. Second,
active-sterile mixing---especially $\nu_e-\nu_s$
mixing---depletes the electron neutrino and antineutrino
populations, reducing the rates of the n $\leftrightarrow$ p
interconversion reactions. Both of these effects cause n/p
to freeze out at a lower temperature.

Using a neutrino ensemble evolution formalism 
\cite{steriles,mckellar} that includes
both neutrino oscillations (with matter effects) and neutrino
collisions, previous authors \cite{steriles} 
have produced exclusion plots
in the $\delta m^2$-$\sin ^2 2\theta$ plane for both 
$\nu_e-\nu_s$ and $\nu_{\mu}-\nu_s$ mixing
\cite{note1}. Here $\delta m^2$
and $\sin ^2 2\theta$ are the difference 
of the squares of the neutrino vacuum mass eigenvalues
and a measure of the vacuum
mixing angle, respectively, associated with two-flavor neutrino mixing.
 These studies
showed that for $\eta > 2.8 \times 10^{-10}$, both the 
$\nu_{\mu}-\nu_s$ solution to the 
atmospheric neutrino problem \cite{atmos}
and the $\nu_e-\nu_s$ large-angle MSW solution to the solar
neutrino problem \cite{solar} are excluded for $Y < 0.247$.

In our study of the primordial D/H dependance of BBN 
constraints on active-sterile neutrino mixing, we have employed 
the same neutrino evolution formalism 
\cite{steriles,mckellar} as previous authors.
We have neglected any net lepton number contributed by 
the neutrinos. (The recently reported effect of active-sterile
neutrino mixing {\em generating} net lepton number does
not occur in the regions of parameter space we consider
here \cite{lepno}.) 
In this case, the neutrino and antineutrino sectors evolve 
identically. A Fermi-Dirac momentum
distribution for all neutrinos is assumed, 
but allowance is made for 
non-equilibrium number densities. The differential equations 
in the formalism yield $n_{\nu_e}, n_{\nu_{\mu}}, n_{\nu_{\tau}}, 
n_{\nu_s}$, and $T$ as functions of time. Here $n_{\nu_x}$ 
denotes the fraction of a full fermionic degree of freedom
contributed by neutrino species $x$, which we shall hereafter
call the ``number density parameter'' of neutrino species $x$.
In the equations below we will take $n_{\nu_x} = n_{\overline
\nu_x}$, since we are working under the assumption that
the net lepton number contributed by the neutrinos is 
negligible.

In our BBN computation
we have employed the Kawano \cite{kawano} 
update of the Wagoner \cite{wagoner} code,
with the latest world average neutron lifetime, $\tau =887.0$ s
\cite{montanet};
the reaction rates of Ref. \cite{smith};
and a correction of $+0.0031$ to $Y$ due to finite nucleon mass
and timestep-dependant effects \cite{kernan}.
We have altered the Kawano code to use the `temperature
series' of neutrino
number density parameters, $n_{\nu_x}(T)$,
to compute the energy density 
contributed by neutrinos and the n $\leftrightarrow$ p 
interconversion rates. The neutrino energy density is 
\begin{equation}
  \rho_{\nu} = {7 \over 8}\, {\pi^2 \over 15} (n_{\nu_e} +
   n_{\nu_{\mu}} + n_{\nu_{\tau}} + n_{\nu_s}) T^4.
\end{equation}
The n $\leftrightarrow$ p rates are
\begin{equation}
  \lambda_{ne \rightarrow p\nu} = K \int_1^{\infty}
    \left(1 \over {1+e^{x z}} \right) \left(1 - {n_{\nu_e} \over
    {1+e^{(x+q)z_{\nu}}} }\right) x (x+q)^2 (x^2-1)^{1/2} \,dx,
\end{equation}
\begin{equation}
  \lambda_{n\nu \rightarrow pe} = K \int_q^{\infty}
    \left(n_{\nu_e} \over
    {1+e^{(x-q)z_{\nu}}} \right)\left(1 \over {1+e^{-x z}} \right)
    x (x-q)^2 (x^2-1)^{1/2} \,dx,
\end{equation}
\begin{equation}
  \lambda_{n \rightarrow pe\nu} = K \int_1^q
    \left(1 \over {1+e^{-x z}} \right) \left(1 - {n_{\nu_e} \over
    {1+e^{(q-x)z_{\nu}}} }\right) x (x-q)^2 (x^2-1)^{1/2} \,dx,
\end{equation}
\begin{equation}
  \lambda_{pe \rightarrow n\nu} = K \int_q^{\infty}
    \left(1 \over {1+e^{x z}} \right) \left(1 - {n_{\nu_e} \over
    {1+e^{(x-q)z_{\nu}}} }\right) x (x-q)^2 (x^2-1)^{1/2} \,dx,
\end{equation}
\begin{equation}
  \lambda_{p\nu \rightarrow ne} = K \int_1^{\infty}
    \left(n_{\nu_e} \over
    {1+e^{(x+q)z_{\nu}}} \right)\left(1 \over {1+e^{-x z}} \right)
    x (x+q)^2 (x^2-1)^{1/2} \,dx,
\end{equation}
\begin{equation}
  \lambda_{pe\nu \rightarrow n} = K \int_1^q
    \left(1 \over {1+e^{x z}} \right)
    \left(n_{\nu_e} \over
    {1+e^{(q-x)z_{\nu}}} \right)x (x-q)^2 (x^2-1)^{1/2} \,dx.
\end{equation}
In these expressions $x\equiv E_e /m_e$, where $E_e$ and $m_e$
are the total electron (or positron) energy and 
rest mass, respectively;
$z \equiv m_e/T$; $z_{\nu}\equiv m_e/T_{\nu}$, where $T_{\nu}$
is the appropriate neutrino temperature; 
$q \equiv \Delta m / m_e$; and $K$
is a constant obtained by solving the equation
\begin{equation}
\left(\lambda_{ne \rightarrow p\nu} +
  \lambda_{n\nu \rightarrow pe} +
  \lambda_{n \rightarrow pe\nu}\right)|_{z\rightarrow \infty}
  = 1/\tau
\end{equation}
for $K$, where $\tau$ is the experimentally 
measured neutron lifetime.

The lower limit on 
$\eta$ obtained from a standard
BBN calculation with $N_{\nu}=3$ is not appropriate for 
BBN with active-sterile mixing. This is because the lower 
limit on $\eta$ depends on the expansion rate, often codified
as an effective number of neutrino generations $N_{\nu}$
\cite{cardall1}.
Since active-sterile mixing increases $N_{\nu}$
(at least for the range of parameter space we consider here
\cite{lepno}), 
it affects the lower bound on $\eta$. 
Therefore, we will plot our results as a function
of the primordial D/H value---the experimentally 
determined quantity---rather than
as a function of $\eta$. These considerations are most 
important for the $\nu_{\mu}-\nu_s$
atmospheric neutrino mixing solution, and much less
important (nearly negligible) for the 
 $\nu_e-\nu_s$ small-angle MSW solution to the solar neutrino
problem.

In Fig. 1, a representative $\nu_{\mu}-\nu_s$
atmospheric neutrino mixing solution
($\delta m^2 = 1.0 \times 10^{-2}$ eV$^2$, $\sin ^2 2\theta=
0.6$ \cite{steriles})
is {\em assumed}, and the resulting BBN $^4$He yield is 
plotted as a function of the BBN D/H yield. The value of $\eta$
(given as $\eta_{10} = 10^{10} \; \eta$)
at various values of D/H is also indicated on the figure. 
For a given value
of D/H, the implied abundance of $^4$He can be interpreted as 
the observational upper limit required to constrain the solution. 
Alternatively, a {\em detection} of these neutrino mixing 
parameters by, for example, 
future atmospheric neutrino experiments would 
yield an independent determination of the primordial $^4$He
abundance, so long as
D/H were known from QAS studies. This could be very 
interesting, given
the recent emphasis on the systematic uncertainties
in the determination of the primordial $^4$He abundance 
as derived from helium recombination lines in
extragalactic HII regions \cite{uncert}. The conclusions reached
from Fig. 1 are essentially the same over the range of $\delta m^2$
($10^{-3}-10^{-1}$ eV$^2$) for the proposed $\nu_{\mu}-\nu_s$
mixing explanation of the atmospheric neutrino problem. 

Fig. 2 is similar to Fig. 1, 
but with a representative  
$\nu_e-\nu_s$ small-angle MSW solution to the solar neutrino
problem assumed ($\delta m^2 = 4.0 \times 10^{-6}$ eV$^2$, 
$\sin ^2 2\theta=8.0 \times 10^{-3}$ \cite{MSWsol}). 
These mixing parameters have only a very
small effect on the BBN $^4$He yield. For the most recent
data from solar neutrino experiments and the standard
solar model \cite{bahcall}, there is no $\nu_e-\nu_s$ 
large-angle MSW solution to the solar neutrino problem
\cite{MSWsol}.
 
Some of the QAS data suggest D/H $\approx 2 \times 10^{-4}$
\cite{hid}.
Figs. 1-2 show the range D/H $=
1.5-2.3 \times 10^{-4}$, as determined in Ref. \cite{rugers}. 
This range of D/H implies a 
lower bound on $\eta$ that is significantly lower than that used
in previous studies. Since a lower $\eta$ implies a lower $^4$He
yield, high D/H relaxes constraints 
on any effect that increases the expansion rate, including
mixing with sterile species \cite{kernan2}. 
Fig. 1 shows, however, that 
for D/H $\approx 2 \times 10^{-4}$, the $\nu_{\mu}-\nu_s$
atmospheric neutrino solution is still somewhat constrained
\cite{note2}
if current observational inferences \cite{olive}
of primordial $^4$He are
correct: $Y = 0.234 \pm 0.003 \pm 0.005$, where the first
error is statistical and the second systematic. Of course,
if this $\nu_{\mu}-\nu_s$
atmospheric neutrino mixing solution were inferred from 
atmospheric neutrino experiments, and QAS studies confirm
D/H $\approx 2 \times 10^{-4}$, the implied $^4$He abundance
of $Y \approx 0.245$ would be significantly higher than the central
value of $Y = 0.234$ cited above.

The $\nu_e-\nu_s$ small-angle MSW solar neutrino solution is 
allowed for D/H $\approx 2 \times 10^{-4}$. Fig. 2 shows that
an observational upper bound of $Y \lesssim 0.232$ would be 
required to restrict this small angle solution if such a high
D/H is indeed the primordial value.

Other very high quality QAS data---arguably better 
\cite{lowd} for the
determination of D/H
than that used in Ref. \cite{hid}---suggest 
D/H $\approx 2 \times 10^{-5}$ \cite{lowd}. 
This value of D/H is incompatible
with standard BBN with $N_{\nu}=3$
\cite{kernan,hata,cardall1} for current observational 
inferences of the primordial $^4$He abundances \cite{olive}, 
and any mixing with
sterile neutrinos would only exacerbate the problem.
As mentioned previously, however, it has been argued that 
$Y$ has been systematically underestimated, and a more
appropriate upper limit on $Y$ may actually be $Y \le 0.255$
\cite{uncert}. It is unlikely
that the systematic error in $Y$ is enough to allow the 
$\nu_{\mu}-\nu_s$
atmospheric neutrino 
solution for D/H $\approx 2 \times 10^{-5}$. 
However, observation of
the $\nu_e-\nu_s$ small-angle MSW solar neutrino solution,
together with a solid determination of 
D/H $\approx 2 \times 10^{-5}$,
would require that $Y$
has been systematically and significantly 
underestimated by about $0.015$ (see Fig. 2),
unless there is non-standard physics during the BBN epoch
\cite{hata}. 
This is a somewhat trivial point, since the mixing parameters
of the small-angle $\nu_e-\nu_s$ MSW solution produce
only very slightly more $^4$He than the standard BBN picture
with $N_{\nu}=3$, for which the ``crisis'' at low D/H is 
well-known \cite{kernan,hata,cardall1}. Useful constraints
on the $\nu_e-\nu_s$ small-angle MSW solar neutrino solution
would require very precise observational knowledge of $\eta$
and $Y$. This
may, however, still be interesting in view of the fact that future
solar neutrino experiments may 
be able to distinguish the sterile neutrino oscillation-based
solution from other solutions \cite{nuexp}. 
Also, many models that seek to
satisfy all available constraints on neutrino properties employ
the $\nu_e-\nu_s$ small-angle MSW solar neutrino solution 
\cite{numod} (but
see Ref. \cite{cardall2}).  

\section*{Acknowledgements}  
This work was supported by the National 
Science Foundation through NSF
Grant PHY-9503384, and by a NASA Theory Grant.


%
%
 \begin{figure}
 \caption{BBN yields for a typical $\nu_{\mu}-\nu_s$ atmospheric 
	neutrino solution ($\delta m^2 = 1.0 \times 10^{-2}$ eV$^2$, 
	$\sin ^2 2\theta=0.6$). The solid curve is the 
	$^4$He mass fraction $Y$ vs. $D/H$. The squares indicate,
	from lower left to upper right,
	$10^{10}\; \eta =$ 1.7, 2.3, 3.0, 4.6, 6.6, 8.6. The dotted lines
	indicate the ranges of ``high'' and ``low'' 
	 D/H inferred from QAS studies. The dashed lines indicate 
	two possible upper limits on
	$Y$: the currently popular $Y=0.245$, and the 
	more conservative $Y=0.255$.}
 \end{figure}

 \begin{figure}
 \caption{BBN yields for a typical $\nu_e-\nu_s$ small-angle MSW
	solution to the solar 
	neutrino problem ($\delta m^2 = 4.0 \times 10^{-6}$ eV$^2$, 
	$\sin ^2 2\theta=8.0 \times 10^{-3}$). The solid curve is the 
	$^4$He mass fraction $Y$ vs. $D/H$. The squares indicate,
	from lower left to upper right,
	$10^{10}\; \eta =$ 1.5, 2.0, 2.6, 3.6, 4.6, 5.6. The dotted lines
	indicate the ranges of ``high'' and ``low'' 
	 D/H inferred from QAS studies. The dashed lines indicate 
	two possible upper limits on
	$Y$: the currently popular $Y=0.245$, and the more conservative
	$Y=0.255$.}
 \end{figure}

%
%


\begin{references}

\bibitem{steriles}X. Shi, D. N. Schramm, and B. D. Fields,
	Phys. Rev. D {\bf 48}, 2563 (1993);
	K. Enqvist, K. Kainulainen, and M. Thomson,
	Nucl. Phys. B {\bf 373}, 498 (1992); and references
	therein.

\bibitem{hid}M. Rugers and C. J. Hogan, report 
	astro-ph/9603084, 1996 (unpublished); M. Rugers
	and C. J. Hogan, Astrophys. J. Lett. {\bf 459}, L1
	(1996); E. J. Wampler {\it et al.}, Astron. Astrophys.
	(to be published); R. F. Carswell {\it et al.}, Mon. Not.
	R. Astron. Soc. {\bf 278}, 506 (1996); A. Songaila,
	L. L. Cowie, C. J. Hogan, and M. Rugers, Nature
	{\bf 368}, 599 (1994); R. F. Carswell {\it et al.},
	Mon. Not. R. Astron. Soc. {\bf 268}, L1 (1994).

\bibitem{lowd}S. Burles and D. Tytler, report astro-ph/9603070,
	1996 (unpublished); D. Tytler, X. Fan, and S. Burles,
	report astro-ph/9603069 (unpublished).

\bibitem{nuexp}P. I. Krastev, S. T. Petcov, and L. Qiuyu,
	IASSNS - AST 96/11, hep-ph/9602033, 1996
	(unpublished); S. M. Bilenky and C. Giunti, 
	Z. Phys. C {\bf 68}, 495 (1995); P. I. Krastev and
	S. T. Petcov, Nucl. Phys. B {\bf 449}, 605 (1995).
	W. Kwong and S. P. Rosen, 
	Phys. Rev. Lett. {\bf 68}, 748 (1992).

\bibitem{numod}G. M. Fuller, J. R. Primack, and Y.-Z.
	Qian, Phys. Rev. D {\bf 52}, 1288 (1995);
	J. J. Gomez-Cadenas and M. C. 
	Gonzales-Garcia, Z. Phys. (to be published);
	S. Goswami, CUPP-95/4, hep-ph/9507212
	(unpublished); 
	E. Ma and P. Roy, Phys. Rev. D
	{\bf 52}, R4780 (1995); E. Ma and J. Pantaleone,
	Phys. Rev. D {\bf 52}, R3763 (1995); R. Foot and
	R. R. Volkas, Phys. Rev. D {\bf 52}, 6595 (1995);
	Z. G. Berezhianai and R. N. Mohapatra, Phys. Rev.
	D {\bf 52}, 6607 (1995);  E. J. Chun, A. S. 
	Joshipura, and A. Y. Smirnov, Phys. Lett. B {\bf 357},
	608 (1995).

\bibitem{kolb}E. W. Kolb and M. S. Turner, {\it The Early
	Universe} (Addison-Wesley, Reading, 1990),
	Chap. 4.

\bibitem{mckellar}B. H. J. McKellar and M. J. Thomson,
	Phys. Rev. D {\bf 49}, 2710 (1994).

\bibitem{lepno}X. Shi, report astro-ph/9602135, 1996 
	(unpublished); R. Foot, M. J. Thomson, and 
	R. R. Volkas, UM-P-95/90, 1995 (unpublished).

\bibitem{note1} $\nu_{\tau}-\nu_{s}$ mixing would have the same 
effects on $Y$
as $\nu_{\mu}-\nu_s$ mixing, but this is of less interest 
since neither the solar nor atmospheric neutrino anomalies 
could be solved with $\nu_{\tau}-\nu_{s}$ mixing.

\bibitem{atmos}T. K. Gaisser, F. Halzen, and T. Stanev,
	Phys. Rep. {\bf 258}, 173 (1995).

\bibitem{solar}W. Haxton, Ann. Rev. Astron. Astrophys.
	{\bf 33}, 459 (1995); J. N. Bahcall, Astrophys. J.
	(to be published).

\bibitem{kawano}L. Kawano, Fermilab-Pub-92/04-A,
	1992 (unpublished).

\bibitem{wagoner}R. V. Wagoner, Astrophys. J. {\bf 179},
	343 (1972); R. V. Wagoner, Astrophys. J. Suppl. Ser.
	{\bf 18}, 247 (1969).

\bibitem{montanet}Particle Data Group, L. Montanet {\it et al.},
	Phys. Rev. D {\bf 50}, 1173 (1994).

\bibitem{smith}M. S. Smith, L. H. Kawano, and R. A. Malaney,
	Astrophys. J. Suppl. Ser. {\bf 85}, 219 (1993).

\bibitem{kernan}P. J. Kernan and L. M. Krauss, Phys. Rev.
	Lett. {\bf 72}, 3309 (1994).

\bibitem{cardall1}C. Y. Cardall and G. M. Fuller, report
	astro-ph/9603071, 1996 (unpublished).

\bibitem{uncert}D. Sasselov and D. Goldwirth, Astrophys.
	J. Lett. {\bf 444}, L5 (1995); C. J. Copi, D. N. Schramm,
	and M. S. Turner, Science {\bf 267}, 192 (1995).

\bibitem{MSWsol}P. I. Krastev and S. T. Petcov in Ref. 
	\cite{nuexp}.

\bibitem{bahcall}J. N. Bahcall and M. Pinsonneault,
	Rev. Mod. Phys. {\bf 64}, 85 (1992).

\bibitem{rugers}M. Rugers and C. J. Hogan in Ref. \cite{hid}.

\bibitem{kernan2}Recently, this point was noted independently
	in P. J. Kernan and S. Sarkar, CWRU-P3-96,
	astro-ph/9603045, 1996 (unpublished).

\bibitem{note2} There is a small but nontrivial
discrepancy between our results and the
those of Shi {\it et al.} \cite{steriles}. 
For $\eta=2.8 \times 10^{-10}$, the
atmospheric neutrino solution would imply $Y = 0.248$
according to Shi {\it et al.} \cite{steriles}, while
we obtain $Y=0.252$. The difference is due to our use of
the $+0.0031$ correction to $Y$, and the fact that we numerically
integrate the altered n$\leftrightarrow$p reactions in our
BBN calculation. Thus the constraints on the $\nu_{\mu}-\nu_s$
atmospheric neutrino solution are not as relaxed as might have
been expected for D/H $\approx 2 \times 10^{-4}$.

\bibitem{olive}K. A. Olive and S. T. Scully, Int. J. Mod. Phys. A
	{\bf 11}, 409 (1996).

\bibitem{hata}N. Hata et al., Phys. Rev. Lett. {\bf 75}, 3977 (1995).

\bibitem{cardall2}C. Y. Cardall and G. M. Fuller, Phys. Rev. D
	(to be published 15 Apr. 1996).

\end{references}
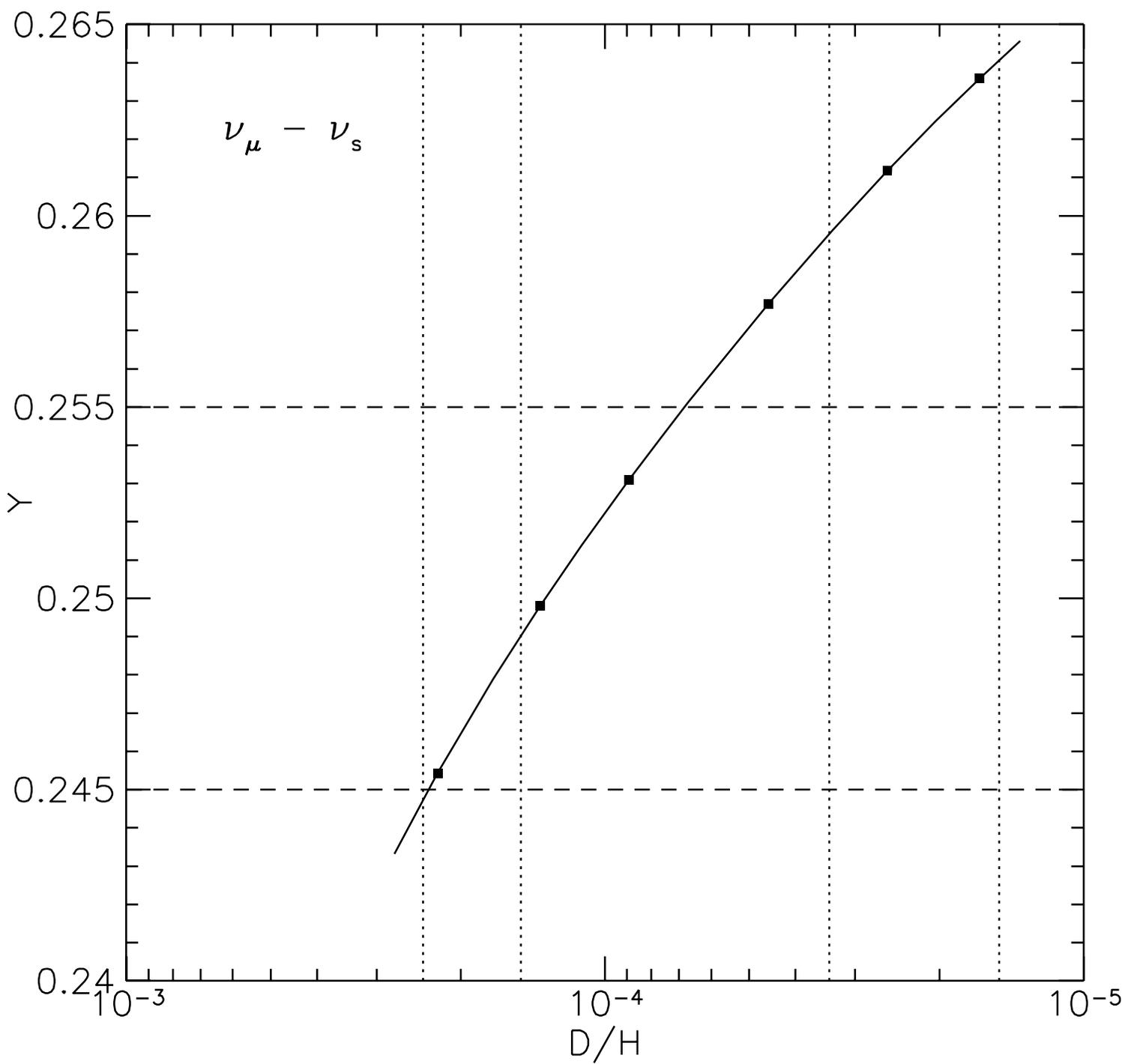
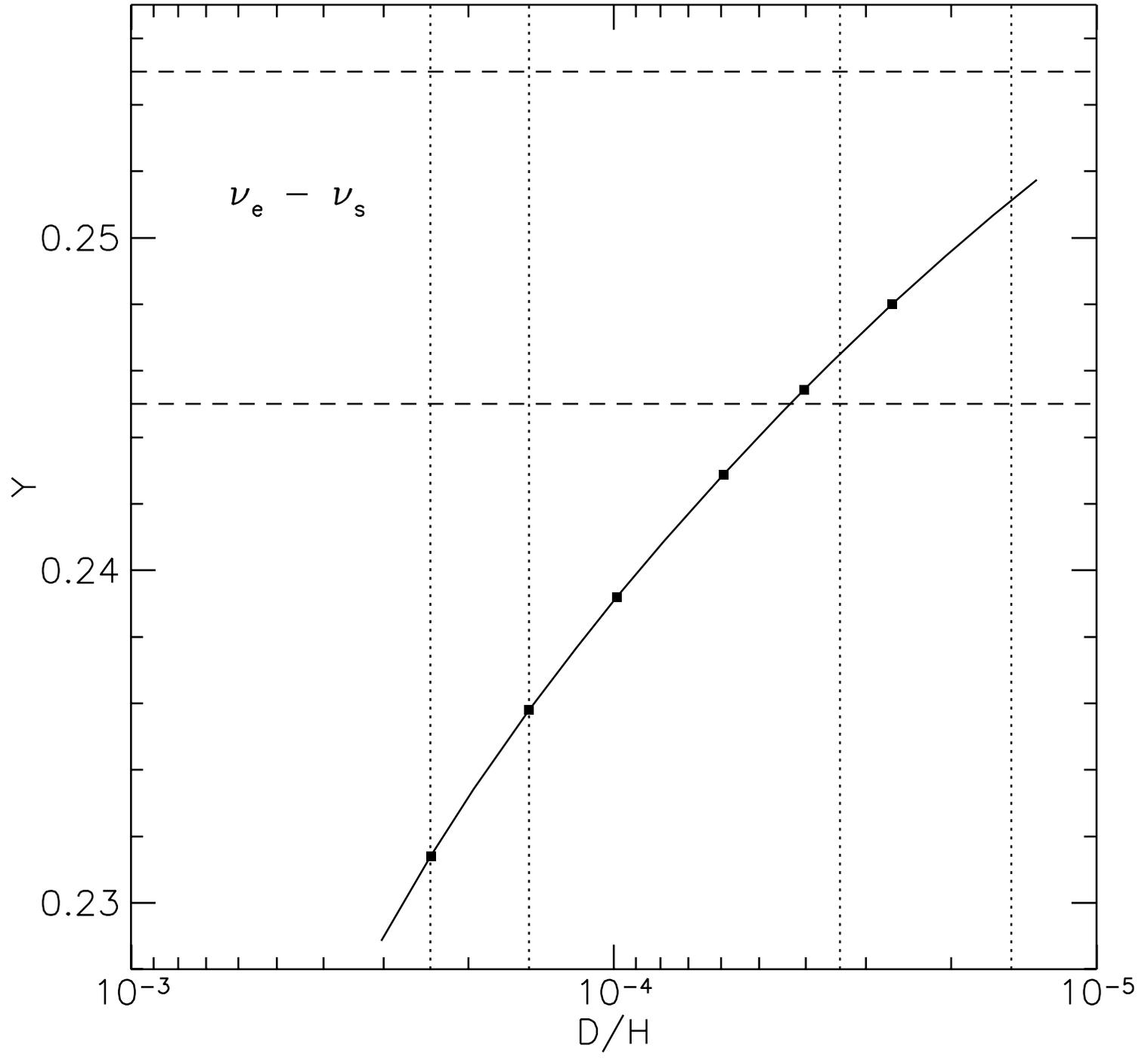
\end{document}